# A State-of-the-Art Review on IoT botnet Attack Detection

Zainab Al-Othman
*Computer Science Department*
*Princess Sumaya University for Technology*
Amman, Jordan
zai20178091@std.psut.edu.jo

Mouhammd Alkasassbeh
*Computer Science Department*
*Princess Sumaya University for Technology*
Amman, Jordan
m.alkasassbeh@psut.edu.jo

Sherenaz AL-Haj Baddar
KASIT, The University of Jordan
*Princess Sumaya University for Technology*
Amman, Jordan
s.baddar@ju.edu.jo sh.alhajbaddar@psut.edu.jo

*Abstract*— The *Internet as we know it Today, comprises several fundamental interrelated networks, among which is the Internet of Things (IoT). Despite their versatility, several IoT devices are vulnerable from a security perspective, which renders them as a favorable target for multiple security breaches, especially botnet attacks. In this study, the conceptual frameworks of IoT botnet attacks will be explored, alongside several machine-learning based botnet detection techniques. This study also analyzes and contrasts several botnet Detection techniques based on the Bot-IoT Dataset; a recent realistic IoT dataset that comprises state-of-the-art IoT botnet attack scenarios.*

## I. Introduction

Undeniably, the Internet of Things has become one of the core components of the Internet, as numerous organizations worldwide regard the seamless integration of IoT devices into their business infrastructure as a top priority. According to recent statistics, the number of connected IoT devices will reach more than 20.4 billion in 2025. It is also expected that the number of active IoT devices will reach nearly 64 billion devices [1]. With the accelerating rate of IoT devices deployment, the security risks, vulnerabilities, and threats will multiply, jeopardizing the integrity and soundness of countless systems worldwide. To further deepen this risk, a substantial proportion of nowadays IoT devices do not adhere to security standards [2]. According to Nokia's threat intelligence report 2019, in 2018, IoT botnet attacks comprised 78% of malware. Due to the wide array of vulnerabilities, many IoT devices, recent and older, form an accessible and attractive target for networks intruders. One of the most occurring attacks that target IoT devices, is the IoT botnet attack, which tries to penetrate and recruit vulnerable IoT devices to create massive networks of "Zombie" or "bot" devices. These networks can be easily controlled remotely by the behind-the-scenes attacker (i.e the "Bot-Master") to generate countless malicious activities. According to the study depicted in [3], botnet attacks are among the most common malicious attacks that target IoT devices. Botnets were the most influential threats that targeted the global Internet and communications ecosystem according to the International Anti-Botnet Guide 2018. Today, due to the ever-evolving botnet networks, a single botnet comprises 30 million bots on its own [4].

Consequently, the existence of effective and responsive solutions to detect botnets has become a necessity. Botnet detection techniques have two main categories; honeynet-based detection techniques, and Intrusion Detection Systems (IDSs). Typically, IDSs can be either signature-based or behavioral, yet hybrid IDSs that combine features from both IDs types have been introduced in the literature [5], [6], [7], [8]. While signature-based IDSs identify bots that match predefined patterns, behavioral IDSs are capable of identifying unseen bots by analyzing network traffic and identifying anomalies pertaining to high network latency, traffic on unusual ports, and high network volume, to mention some [9]. Anomaly-based detection techniques are further classified into host-based and network-base. In host-based IDSs, detection mainly focuses on analyzing the behavior of individual machines, while network-based, detection focuses on analyzing the collective network traffic [10]. Several approaches to design IDSs have been explored, among which are machine-learning techniques, and more recently deep learning techniques. Studies on applying machine-learning techniques to botnet detection promise different degrees of success. Thus, we explore and contrast recent studies on applying machine-learning techniques to IoT botnet attacks detection.

## II. State-of-the-Art on Botnet Detection

This section reviews and contrasts recent techniques that have been proposed to address botnet detection, using the Bot-IoT dataset in particular. The Bot-IoT dataset is a recent realistic dataset that comprises traces of multiple attacks carried out using IoT bots.

A. *Botnet Detection Techniques Using the Bot-IoT Dataset*

Due to the complexity of the nature of anomaly detection, there has been only limited success in intrusion detection proposed in the literature, including but not limited to, botnet attacks detection. In this part of the study, we depict and discuss recent techniques reported in the literature to help detect botnet attacks using the Bot-IoT dataset.

Koroniotis et al. applied three machine-learning techniques; namely Support Vector Machine (SVM), Recurrent Neural Network (RNN), and Long-Short Term Memory Recurrent Neural Network (LSTM-RNN) to the Bot-IoT subdataset, with all 46 features [11]. The experiment results showed that the SVM classifier took the longest training time, relatively speaking when using all the features, but with the highest accuracy and recall rates. Using only the 10 highlighted features, this classifier achieved the highest precision and lowest fall-out rates, relatively speaking. In other words, the SVM achieved the best outcomes compared to the two other techniques.

Several studies proposed signature-based IDSs for detecting IoT attacks, like the one depicted in [12], where a model that generated attack signature rules was proposed. The model utilized a J48 decision tree, and used almost 660,000 instances from the Bot-IoT dataset, with 43 features. The model generated 24 rules, 16 of which pertained to attack traffic; also, attack rules comprised one "Theft" rule, one "DDoS" rule, and all remaining rules were for "Probing". This model augments public signature-based IDS systems such as Snort and Suricata to detect IoT attacks. Moreover, it offers a fast and light IDS for known-attacks that would fit the nature of IoT devices should it be hosted on the devices themselves. Nevertheless, both performance metric and evaluation results were not included in this study.

The study presented in [5] introduced a hybrid Intrusion Detection System (HIDS) that aimed at improving the accuracy of detecting IoT attacks. This system combines the Signature-based IDS for well-known attacks detection, with a behavioral IDS for zero-day attacks detection. Classification results from both systems were combined using the Boosting method, where the Signature-based part comprised a C5 decision tree, whereas the behavioral part comprised a One-Class SVM, with only 13 out of the original 46 features from the Bot-IoT datasets while performing binary classification. The accuracy of detection, using only the Signature-based part, was 93.30%, while the accuracy with using only the behavioral part was 92.50%. On the other hand, the combined accuracy of both reached 99.97%. The accuracy of the detection of the proposed system was compared with the accuracy of other algorithms such as C4.5, Naïve Bayes (NB), Random Forest (RF), multi-layer perceptron network, Support Vector Machine (SVM), Classification And Regression Tree (CART), and K-Nearest Neighbor (KNN) algorithms. As their results showed, the proposed system achieved the highest accuracy rate, and the Random Forest algorithm showed the second-highest accuracy rate of about 92.67%. The authors claim their approach resulted also in the least false alarm rate, compared to the other classification algorithms, yet, required longer time to train the models.

Baig et al proposed a solution to the DoS attacks that target IoT wireless sensor networks (WSNs) [13], in which two novel frameworks were proposed, namely, Multi-Scheme and Voting, that combine the averaged of one-dependence estimator (A1DE), and two-dependence estimator (A2DE). As for the Multi-Scheme mechanism, either A1DE or A2DE classifiers were selected for the testing phase, based on these classifiers' accuracy in the training phase using the entire dataset. The Voting scheme, however, combined the outputs of the two selected classifiers, A1DE, and A2DE by applying the average of probabilities. More precisely, if both classifiers produced the same results for a given sample of data, then the result is accepted during testing. However, if there is a gap between both classifiers' outcomes, then the result is randomly chosen. Those frameworks were compared with other classifiers, including A1DE, A2DE, Naïve Bayes, Bayesian Network, C4.5, and MLP classifiers. This study also utilized the Bot–IoT dataset to help calculate the detection accuracy and computational time, where only 477 normal samples and 3,668,045 attack samples were used. The experiments showed that, when using only 5 of the original features, the performance of the Multi-Scheme classifier and A2DE classifier was approximately the same, in terms of detection accuracy. However, the Multi-Scheme classifier took significantly longer to train compared to the A2DE classifier. Although the Multi-Scheme classifier also took longer to train compared to the Voting classifier, its accuracy results were better. The authors also evaluated both the A1DE and A2DE classifiers and compared their results with the performance results of the radial neural network (RNN) that was reported in [11]. The results showed that the A2DE classifier produces the best accuracy for both DoS-HTTP and DDoS-HTTP attacks, compared to the performance of the RNN. The accuracy of the RNN was close to 98%, while the accuracy of A1DE, and A2DE, was close to 99.8% for DoS-HTTP attacks. Although the results showed that the A2DE classifier was better than the RNN classifier, it only focused on the DoS attack types. Other botnet attack scenarios, such as OS Fingerprinting, Service Scan, Data exfiltration, and keylogging, were ignored.

One technique that is featured in recent solutions to botnet detection is deep learning, which while has proven to be useful in detecting and mitigating attacks targeting IoT networks, also faces several challenges. Next, we highlight sample studies that utilized deep learning in IoT botnet detection.

Several studies explored the challenges that face deep learning algorithms in security applications; the study depicted in [14] for example, measured the impact of the adversarial attacks on an IoT network with an IDS that utilizes machine learning. Adversarial attacks are attempted via instances with intentional feature perturbations, designed by an attacker to force the model to make wrong predictions. The adversarial samples, in this study, were synthesized by the authors with the use of the Adversarial Robustness Toolbox (ART) framework [15]. Three methods were used to generate these malicious instances, the first of which was using the Fast Gradient Sign Method (FGSM). This method implements a one-step gradient update across the gradient sign direction for each instance in the dataset [16]. The second method, on the other hand, utilized the Basic Iteration method (BIM), which optimizes FGSM [17]. The third method was the Projected Gradient Descent (PGD), which is very similar to FGSM, but neglects its random start feature [18]. To detect adversarial attacks, two deep learning classifiers were evaluated; Feed-forward Neural Network (FNN) and Self-normalizing Neural Networks (SNN). To assess their classification accuracy, the Bot-IoT subdataset was used. Experiments showed that, when using the adversarial-free dataset, the FNN classifier produced the best accuracy which exceeded 94%. However, when the adversarial samples were added, the SNN algorithm showed better flexibility compared to the FNN classifier. Lastly, when applying feature normalization, both algorithms showed higher accuracy with the adversarial-free dataset, which emphasizes the importance of normalization. Both deep learning classifiers were found to be susceptible to adversarial samples.

Another study that proposed using a deep learning-based intrusion detection framework for IoT networks is depicted in [19], where an FNN classifier was utilized. The accuracy rate of binary classification topped 0.99 for DDoS/DoS and reconnaissance attacks, while it was above 0.92 with the information theft attacks. In the case of the multi-class classification, the detection accuracy of the DDoS/DoS was 99.414%, and it was 98.375% with the reconnaissance attacks, while it was 88.918% with Information theft attacks. In the case of the DoS over UDP attacks, the classifier predicted 0.1% of the malicious packets as normal packets. Also, for the Reconnaissance-Service-Scanner attacks, the classifier predicted 0.6% of the normal packets as malicious packets and 0.3% of the malicious packets as normal packets. It is worth noting that the detection accuracy of the information theft attacks, i.e. the data exfiltration and the keylogging attacks, were lesser in both binary and multi-class classification, compared to the rest of the attacks. In the case of the binary classification, the detection accuracy of the data exfiltration attacks was 92.78%, and 96.82% for the keylogging attacks. In the case of the multi-class classification, the detection accuracy for the information theft attacks was 88.918%. This result could be attributed to the fact that the Bot-IoT dataset is imbalanced; the information theft attacks instances comprise only 0.002% of the original dataset.

Seven deep learning algorithms for intrusion detection were evaluated in [20]. These algorithms were grouped in two categories; the first utilized deep discriminative models, which comprised Recurrent Neural Networks (RNN), Deep Neural Networks (DNN), and Convolutional Neural Networks (CNN). The second category comprised generative/unsupervised algorithms, which spanned restricted Boltzmann machines (RBM), Deep Belief Networks (DBN), a Deep Boltzmann machine (DBM), and a Deep Auto-encoder (DA). This study used two realistic traffic datasets, the CSE-CIC-IDS2018 dataset [11], and the Bot-IoT dataset [21] to assess the performance of the examined deep learning algorithms. The CSE-CIC-IDS2018 dataset comprises about 15.5 million records. The outcomes showed that the CNN algorithm achieved best accuracy in binary classification and the training time results for both datasets, with 100 hidden nodes and a learning rate of 0.5. For the CSE-CIC-IDS2018 dataset, the accuracy reached 97.376%, and the training time was almost 330 seconds, while for the Bot-IoT dataset the accuracy reached 98.371%, with almost 1367 seconds for training.

Artificial Neural Networks (ANN) were utilized in [22] to detect DDoS attacks. The Bot-IoT dataset is imbalanced, as described earlier, thus, the Synthetic Minority Over-sampling Technique (SMOTE) was used to increase the number of normal samples so that they match the size of DDoS records. The number of legitimate traffic records became almost 1.3 million instances for training, and almost

656,000 instances for testing. The proposed system is for binary classification, and was trained using 66% of the dataset and tested by the remaining 34%. Only 41 features were used from the original set 46 features. The experimental results showed that, when using the SMOTE technique, the detection accuracy for the DDoS attack was up to 100%.

The study depicted in [23] proposed a framework for the intrusion detection in smart home IoT networks that utilized both optimization and deep learning techniques. The proposed framework comprised 3 major processes; flow extraction, feature selection, and detection. In the flow extraction process, several tools including Wireshark, tcpdump, and Ettercap, were used to collect traffic traces, while the Bro and Argus tools were used to extract flows from the aggregated traffic. The resulting flows were fed to the feature selection process, which applied a Particle Swarm Optimization (PSO) algorithm to select the best features for classification. In the last process, a Deep Neural Network (DNN) was built using the outcomes from the first two processes to trace and detect IoT attacks. To assess the performance of the proposed framework, the Bot-IoT subdataset and the UNSW-NB15 dataset were used [24]. The Bot-IoT subdataset was split into two portions; 80% for training and 20% for testing. Three types of deep networks were designed for the purpose of experimentation; a plain neural network with the feature selection process by-passed, an MLP network where the first 13 features were combined into one, and an MLP network using only 13 of the features from the original data set. The plain neural network achieved high detection accuracy, but with high False Positive Rate (FPR). The MLP network with the combined features achieved high detection accuracy almost equal to 95%, with lower FPR and FNR values compared to the plain network. The last MLP network achieved the highest detection accuracy with near zero FPR and False Negative Rate (FNR) ratios.

The IoT networks botnet IDs proposed in [25] utilized Fuzzy Rule Interpolation (FRI), where only 50K instances with the top-five features from the original Bot-IoT dataset were extracted to help train the proposed system. The proposed system was designed using sparse fuzzy model identification (RBE-DSS). Besides, the Least Squares based Fuzzy Rule Interpolation (LESFRI) technique was used to implement its inference engine. As the experimentation results showed, the proposed system achieved a reasonable detection rate equal to 95.4%. It was also noted that even when none of the fuzzy rules explicitly matched the attack knowledge base, the proposed system still had the potential to generate the necessary warnings.

B. *Summary and Discussion*

This subsection depicts a summary of the studies presented in this review, highlighting their most important aspects. By examining the studies depicted in this review, we notice the following common trends:
- Most IoT-Botnet detection solutions opt for a behavioral approach that is mainly based on machine learning, with several state-of-the-art studies deploying deep learning. Hybrid and signature-based approaches are not common in state-of-the-art solutions.
- Most state-of-the-art solutions utilize the Bot-IoT dataset, however, only few studies addressed its imbalanced nature. Most observed studies simply glossed-over under-represented attack categories.
- Most studies considered the accuracy rate the main performance indicator of their designated classification models.
- Most studies opted for one family of classification algorithms when they designed their models.
- Few studies shed enough light on their feature selection approaches or utilized mainly a single feature selection technique in their model.

These trends highlight some issues that need to be addressed when designing IoT-Botnet detection systems:
- The lack of a feedback system that would help the detection model learn from its mistakes and improve their chances at detecting zero-day attacks.
- The absence of a systematic approach for handling the imbalance of the Bot-IoT dataset, implies a false sense of accuracy detection.
- Even if the imbalanced nature of the Bot-IoT dataset was addressed when a given detection model is designed, accuracy rate is not necessarily enough for producing a true assessment of the effectiveness of the model; other performance metrics, especially False Positive Rate, are crucial for determining the quality of the detection model.
- Contrasting and/or combining more than one type of machine learning classification algorithms may produce more sensitive, and thus better, detection models.
- Feature selection in the Bot-IoT dataset needs further study; the original set of features is large, and different feature selection techniques can provide different perspectives to the classification process, which would help improve the detection accuracy. Also, contrasting the cost of searching for a suitable subset of the features to the performance of the detection models needs to be evaluated.

Thus, this study recommends the deployment of the hybrid approach for botnet IoT detection, as a hybrid IDS that combines signature-based and behavioral detection may stand a better chance at detection IoT botnets in a timely fashion. In such a system, the signature-based detection would help quickly identify previously-seen attacks, while the behavioral detection component would address remaining zero-day/unrecognizable attacks. This implies that a feedback subsystem must be incorporated with such IDS to help it "patternize" newly-detected attacks and add them to the pattern-library that would be associated with the signature-based component of this IDS.

## III. CONCLUSION

IoT devices and networks have become an integral part of the Internet, yet they suffer from security loopholes and vulnerabilities. Security has not been a design goal in most widely-used IoT devices. Thus, several recent attacks managed to penetrate these devices and recruit them to perform severe attacks.

This review sheds light on the state-of-the-art solutions that address IoT botnet detection utilizing the Bot-IoT dataset, and with an emphasis on botnet attacks. It categorizes the proposed solutions and depicts their main aspects, goals, and results. Moreover, a further study of the depicted solutions was performed, where the main trends in these solutions were highlighted, alongside the set of issues exhibited in them. Finally, this study recommends a set of approaches to be deployed in future IoT botnet detection solutions that would help alleviate the issues current solutions have.